\documentclass[amsmath,amssymb,nofootinbib,preprint]{revtex4}

\usepackage{latexsym,comment}
\usepackage{amssymb}
\usepackage{amsfonts}
\usepackage{amsmath,color}
\usepackage{graphicx,epsfig}
\usepackage{bm}

\def\ber{\begin{eqnarray}}
\def\eer{\end{eqnarray}}
\def\beq{\begin{equation}}
\def\eeq{\end{equation}}

\begin{document}

\title{Constraining  $f(T)$ gravity in the Solar System}

\author{Lorenzo Iorio}
\email{lorenzo.iorio@libero.it}
\affiliation{Ministero dell'Istruzione dell' Universit\`{a}  e della Ricerca (M.I.U.R)\\
 Fellow of the Royal Astronomical Society (F.R.A.S.), Viale Unit\`{a} di Italia 68, 70125 Bari (BA), Italy}

\author{Ninfa Radicella}
\email{ninfa.radicella@sa.infn.it}
 \affiliation{Dipartimento di Fisica �E.R. Caianiello�, Universit\`a di Salerno, Via Giovanni Paolo II 132,  Fisciano (Sa), Italy\\
 INFN, Sezione di Napoli, Gruppo Collegato di Salerno,  Napoli, Italy}

\author{Matteo Luca Ruggiero}
\email{matteo.ruggiero@polito.it}
\affiliation{DISAT, Politecnico di Torino, Corso Duca degli Abruzzi 24,  Torino, Italy\\
 INFN, Sezione di Torino, Via Pietro Giuria 1, Torino, Italy}

\date{\today}

\begin{abstract}
In the framework of $f(T)$ theories of gravity, we solve the field equations for  $f(T)=T+\alpha T^{n}$ in the weak-field approximation and for spherical symmetry spacetime.  Since $f(T)=T$ corresponds to Teleparallel Gravity, which is equivalent to General Relativity, the  non linearity of the Lagrangian are expected to produce perturbations of the general relativistic solutions, parameterized by $\alpha$. Hence, we use the $f(T)$ solutions to model the gravitational field of the Sun and exploit data from accurate radio-tracking of spacecrafts orbiting Mercury and Saturn to infer preliminary bounds on the model parameter $\alpha$ and on the cosmological constant $\Lambda$.
\end{abstract}

\maketitle

\section{Introduction}\label{sec:intro}

The geometrical structure of Teleparallel Gravity (TEGR) \cite{pereira,Aldrovandi:2003xu,Maluf:2013gaa} is constituted by a Riemann-Cartan space-time, endowed with the Weitzenb\"ock connection; the latter, unlike the Levi-Civita's one, is not symmetric and gives rise to torsion but it is curvature-free. On the other hand, the space-time of General Relativity (GR) is endowed with the symmetric Levi-Civita connection, which is torsion-free and gives rise to the Riemannian curvature.
Space-time curvature defines the gravitational interaction in GR: the field equations are obtained from the Einstein-Hilbert Lagrangian, which contains  the curvature scalar $R$. By contrast, in TEGR  torsion plays the role of curvature, and the field equations are obtained from a Lagrangian containing the torsion scalar $T$.  Even though the two theories have a different geometrical structure,  they share the same dynamics: this means that  every solution of GR is also solution of TEGR.

The interest in gravity theories with torsion dates back to Einstein: in fact, in  his attempt to formulate a unified theory of gravitation and electromagnetism,  Einstein made use of tetrads and torsion\cite{einstein1,einstein2,einstein3}. There is another important approach in which torsion is used to determine the geometry of space-time, the so-called Einstein-Cartan Theory (ECT)  of gravity: from a geometrical viewpoint, ECT is an extension of GR to Riemann-Cartan spaces, which have both curvature and torsion; in particular,  torsion is related to the density of spin \cite{Hehl:1976kj,klein,Ruggiero:2003tw}. Actually,  the effects of spin and torsion are significant only at very high densities of matter: as a consequence, ECT  is a viable theory of gravity since its observational predictions are in perfect agreement with the
classical tests of GR; moreover, it is suggested that ECT will prove to be a better classical limit of a future theory of quantum gravity \cite{trautmann}.

In Teleparallel Gravity the tetrads field plays the role of the dynamical field instead of the metric field.  Recently \cite{Bini:2015xqa},  the properties of  Weitzenb\"ock connection have been studied with reference to the  nonlocal generalization of GR~\cite{HM1, HM2, RM, Mas}. The so-called  $f(T)$ theories are a generalization of TEGR, in which the gravitational Lagrangian is an  analytic function of the torsion scalar.  Actually, these theories are not equivalent to GR  \cite{Ferraro:2008ey, Fiorini:2009ux}: this is why they have been considered in cosmology as candidates to solve the cosmic acceleration puzzle\cite{cardone12,Myrzakulov:2010vz,Nashed:2014lva,Yang:2010hw,bengo,kazu11,Karami:2013rda,sari11,cai11,capoz11,Bamba:2013jqa,Camera:2013bwa}. Beyond cosmology, other aspects of $f(T)$ theories have been investigated, such as exact solutions, in lower or higher dimensional space-time, also in presence of the electromagnetic field  \cite{Wang:2011xf,Ferraro:2011ks,Gonzalez:2011dr,Capozziello:2012zj,Rodrigues:2013ifa,Nashed:uja,Nashed:2015qza,Junior:2015fya,Bejarano:2014bca,ss3,ss4,ss6,ss7}.

\textcolor{black}{It is important to emphasize that $f(T)$ have additional degrees of freedom \cite{li11,Li:2010cg} with respect to TEGR (and, hence, to GR), that are exploited  in the aforementioned cosmological applications; however, these degrees of freedom do not contribute to the gravitational radiation when considering the first order perturbation theory, as shown in \cite{Bamba:2013ooa}.}

 They are related to the fact that the equations of motion are not invariant under local Lorentz transformations: as a consequence, there exists a preferential global reference frame defined by the autoparallel curves of the manifold that solve the equations of motion. So, even though the form of the space-time metric can be obtained on the basis of the physical symmetries, it is necessary to pay attention to correctly define the form of tetrad field. In particular, as clearly discussed  in  \cite{tamanini12},  a diagonal tetrad  is not the best choice for  non-flat homogenous and isotropic cosmologies (Friedman-Lemaitre-Robertson-Walker universes) and for spherically symmetric space-times (Schwarzschild or Schwarzschild-de Sitter solutions). Spherically symmetric solutions in $f(T)$ gravity are important also because these solutions, describing the gravitational field of point-like sources, can be used to constrain these theories with planetary motions in the Solar System. A weak-field solution, suitable to model the gravitational interaction in the Solar System,  was obtained in  \cite{Iorio12} for a Lagrangian in the form $f(T)=T+\alpha T^{2}$, using a diagonal tetrad; $\alpha$ is a small constant which parameterizes the departure from GR. Observations of the rates of change of perihelia allowed to constrain the theory parameter $\alpha$; tighter constraints on $\alpha$ were obtained in  \cite{Xie:2013vua}.  \textcolor{black}{It is interesting to point out that such a form of Lagrangian can be found also in Born-Infeld gravity \cite{Fiorini:2013kba}, and its cosmological implications have been investigated in \cite{Jana:2014aca}.}

In a recent paper \cite{Ruggiero:2015oka} we studied spherically symmetric solutions, in the weak-field approximation, for Lagrangians in the form  $f(T)=T+\alpha T^{n}$, with $|n| \neq 1$,   and solved the field equations using a non diagonal tetrad. We showed that, to lowest approximation order, the perturbations of the corresponding GR solutions (Schwarzchild or Schwarzchild- de Sitter) are in the form of power laws $\propto \alpha\  r^{2-2n}$, and we estimated the $\alpha$ parameter for $n=2$.

In this paper, we use the solutions found in \cite{Ruggiero:2015oka} to obtain new bounds on $\alpha$ and on the cosmological constant $\Lambda$, by exploiting data from tracking of spacecraft orbiting  some of the major bodies of the Solar System like MESSENGER (Mercury) and Cassini (Saturn).

This work is organized as follows: in Section \ref{sec:fT-gravity} we briefly review the theoretical framework of $f(T)$ gravity and the field equations and, then, we give the  solutions for spherically symmetric space-times, in weak-field approximation; in Section \ref{sec:bound} we    infer bound on the model parameters; eventually, summary and conclusions are Section \ref{sec:conc}.

\section{Weak-field spherically symmetric solutions for $f(T)$ gravity}\label{sec:fT-gravity}

In $f(T)$ gravity the metric tensor can be viewed as a subsidiary field, and the tetrad  field is the dynamical object. Given a coordinate basis, the components of the tetrad field  $e^a_\mu$   are related to the metric tensor by
\beq
g_{\mu \nu}(x) = \eta_{a b} e^a_\mu(x) e^b_\nu(x) \ ,  \label{eq:gmunueta}
\eeq
where $\eta_{a b} = \text{diag}(1,-1,-1,-1)$. Notice that latin indexes refer to the tangent space while greek indexes label coordinates on the manifold.
The field equations are obtained from the action\footnote{We use units such as $c=1$.}

\begin{equation}
{\cal{S}} = \frac{1}{16 \pi G} \int{ f(T)\, e \, d^4x} + {\cal{S}}_M \ ,
\label{eq: action}
\end{equation}
where $e = \text{det} \  e^a_\mu = \sqrt{-\text{det}(g_{\mu \nu})}$ and ${\cal{S}}_M$ is the action for the matter fields. In Eq. (\ref{eq: action}) $f(T)$ is a differentiable function of the \textit{torsion scalar} $T$, which is defined as
\beq
T = S^\rho_{\ \mu \nu} T_\rho ^{\ \mu \nu} \ , \label{eq:deftorsions}
\eeq
where the \textit{contorsion tensor} $S^\rho_{\ \mu \nu}$ is defined by
\beq
S^\rho_{\ \mu \nu} = \frac{1}{4} \left ( T^{\rho}_{\ \ \mu \nu} - T_{\mu \nu}^{\ \ \rho}+T_{\nu \mu}^{\ \ \rho} \right ) +
\frac{1}{2} \delta^\rho_\mu T_{\sigma \nu}^{\ \ \sigma} - \frac{1}{2} \delta^\rho_\nu T_{\sigma \mu}^{\ \ \sigma} \ , \label{eq:defcontorsion}
\eeq
and the \textit{torsion tensor} $T^\lambda_{\ \mu \nu} $ is
\beq
T^\lambda_{\ \mu \nu} = e^\lambda_a \left( \partial_\nu e^a_\mu - \partial_\mu e^a_\nu \right ) \ . \label{eq:deftorsiont}
\eeq
The variation of the action with respect to $e^a_\mu(x)$ gives the following field equations:
\beq
e^{-1}\partial_\mu(e\  e_a^{\ \rho}   S_{\rho}^{\ \mu\nu})f_T+e_{a}^{\ \lambda} S_{\rho}^{\ \nu\mu} T^{\rho}_{\ \mu\lambda} f_T
+  e_a^{\ \rho}  S_{\rho}^{\ \mu\nu}\partial_\mu (T) f_{TT}+\frac{1}{4}e_a^{\nu} f = 4\pi G e_a^{\ \mu} {\cal{T}}_\mu^\nu,
\label{eq: fieldeqs}
\eeq
where ${\cal{T}}^\nu_\mu$ is the matter energy\,-\,momentum tensor and subscripts $T$ denote differentiation with respect to $T$.

We are interested in spherically symmetric solutions of the field equations, so we write the space-time metric in the general form
\begin{equation}
ds^2=e^{A(r)}dt^2-e^{B(r)}dr^2-r^2 d\Omega^2 \ , \label{metric}
\end{equation}
where $d\Omega^{2}= d\theta^2+ \sin^2 \theta d\phi^2$. We
follow the approach described in \cite{tamanini12} to derive the field equations
for the non diagonal tetrad:
\beq
e_\mu^a=\left( \begin{array}{cccc}
e^{A/2}         &   0                                             &   0                                            &    0         \\
0                 &e^{B/2} \sin{\theta}\cos{\phi}   & e^{B/2} \sin{\theta}\sin{\phi}&  e^{B/2} \cos{\theta}\\
0                 &-r \cos{\theta}\cos{\phi}   & -r  \cos{\theta}\sin{\phi}&  r \sin{\theta}\\
0                 &r \sin{\theta}\sin{\phi}   & -r  \sin{\theta}\cos{\phi}&  0\\
 \end{array} \right) \label{eq:ndtetrad}
\eeq
We do emphasize that, because of the lack of local Lorentz invariance, tetrads connected by local Lorentz transformations lead to the same metric - i.e. the same causal structure - but different equations of motions, thus physically inequivalent solutions. In particular, even in an extremely symmetric case like the static spherically symmetric one, a diagonal tetrad that gives back the metric in eq. (\ref{metric}) is not a good choice since the equations of motion for such a choice would constrain a priori the form of the Lagrangian. Actually, this is not the case for the non-diagonal tetrad chosen in eq. (\ref{eq:ndtetrad}). In this case, the  torsion scalar turns out to be
\begin{equation}\label{torsionscalar}
T=\frac{2e^{-B(r)}(1+e^{B(r)/2})}{r^{2}}\left[1+e^{B(r)/2}+r A'(r)\right].
\end{equation}
Exact solutions in vacuum  and in presence of a cosmological constant of the above field equations (\ref{eq: fieldeqs}) are  discussed in \cite{tamanini12} in details. If a vacuum solution of $f(T)$ has $T=0$ it will be a solution of TEGR (and hence GR) as well \cite{Ferraro:2011ks,Bejarano:2014bca}. In fact, the equations of motion with $T=0$ reduce to the Einstein equations of motion with a cosmological constant $\Lambda=\frac{f(0)}{2 f'(0)}$ and one can even avoid the Lambda term by choosing $f(0)$=0 and $f'(0)\neq0$.  Here, we are interested in weak-field solutions with non constant torsion scalar, i.e. $T'=dT/dr \neq 0$: as shown in \cite{tamanini12}, in this case the most general vacuum solution is different from the Scwharzschild - de Sitter one. This is interesting because new features arise with respect to the vacuum solutions of GR, so deviations from GR can be in principle observed.

In the weak-field limit, that is suitable to describe the gravitational field in the solar system, we may write $\displaystyle  e^{A(r)}=1+A(r), \quad e^{B(r)}=1+B(r)$; furthermore, we  confine ourselves to linear perturbations. As shown in \cite{Ruggiero:2015oka}, if we consider Lagrangians in the form $f(T)=T+\alpha T^n$, where $\alpha$ is a small constant, parameterizing the departure of these theories from $GR$, and $|n| \neq 1$, in presence of a cosmological constant $\Lambda$, we obtain the following solutions
\beq
A(r)=-\frac{2GM}{r}-\alpha \frac{r^{2-2n}}{2n-3}2^{3n-1} -\frac 1 3 \Lambda r^{2}
\label{eq:solakn}
\eeq

\beq
B(r)=\frac{2GM}{r}+\alpha \frac{r^{2-2n}}{2n-3} 2^{3n-1} \left(-3n+1+2n^{2} \right)+\frac 1 3 \Lambda r^{2}
\label{eq:solbkn}
\eeq
for the gravitational field around a point-like source $M$.  In particular, if $\Lambda=0$, we obtain vacuum solutions.  We see that the weak-field limit of  Schwarzschild - de Sitter solution is perturbed by terms that are proportional to $\alpha$. Moreover, the torsion scalar is

\beq
T(r)=\frac{8}{r^{2}}+2\alpha r^{-2n}2^{3n} \left(n+1 \right) \label{eq:torsionnk}
\eeq

For $n=2$, the space-time metric is
\beq
ds^2=\left(1-\frac{2GM}{r} -32 \frac{\alpha}{r^{2}}\right)dt^2-\left(1+ \frac{2GM}{r} +96 \frac{\alpha}{r^{2}} \right)dr^2-r^2 d\Omega^2 \label{eq:metricasol1}
\eeq
Notice that the perturbations due to the non linearity of the Lagrangian have the same functional behaviour as in \cite{Iorio12}, but different numerical coefficients: this is not surprising, since the field equations are not the same.

On writing the perturbation terms due to the deviation from GR in the form
\beq
A_{\alpha}(r)=\ \alpha a_{n} r^{2-2n}, \quad B_{\alpha}(r)= \alpha b_{n} r^{2-2n}  \label{eq:defABa}
\eeq
where $\displaystyle a_{n}= \frac{2^{3n-1}}{2n-3}, \, b_{n}=\frac{2^{3n-1}}{2n-3} \left(2n^{2}-3n+1 \right)$, we see that  they go to zero both when $r \rightarrow \infty$ with $n > 1$ and  when $r \rightarrow 0$ with $n<1$. In the latter case,  in order the keep the perturbative approach self-consistent, a maximum value of $r$ must be defined  to consider these terms as perturbations of the flat space-time background.

Eventually, we emphasize that our approach can be applied to an arbitrary polynomial correction $p(T)$ to the torsion scalar, i.e. to $f(T)=T+p(T)$: in particular,  by writing an arbitrary function as a suitable power series, it is possible to evaluate its impact  as a perturbation of the weak-field spherically symmetric solution in GR: the n-th term of the series gives a contribution proportional to $r^{2-2n}$.

\section{Phenomenological bounds on $\alpha$ and $\Lambda$ from Solar System's orbital motions}\label{bounds_from_ranging} \label{sec:bound}

As we have seen above, for Lagrangians in the form $f(T)=T+\alpha T^n$ the perturbations to the GR solution are in the form $\displaystyle  \simeq \alpha  r^{2-2n}$. We focus on $n>1$: actually, since realistic deviations from GR are expected to be small in the Solar System, we can consider, to a first approximation, the solution for $n=2$. Moreover, we do include the contribution due the cosmological constant $\Lambda$: in other words, we consider a space-time metric in the form (\ref{eq:metricasol1}). Hence, here we will phenomenologically infer preliminary bounds on $\alpha, \Lambda$ from the latest results in the field of the Solar System planetary ephemerides \cite{Mess_range, Fienga_Mess, Cass_VLBI, Cass_range}. In doing so, we will suppose that orbital motions follow metric geodesics. Indeed, it is useful to remember that in theories with torsion there is a  distinction between the test particles trajectories: {autoparallels,} or affine geodesics, are curves along which the velocity vector is transported parallel to itself, by the space-time connection; {extremals,} or metric geodesics, are  curves of extremal space-time interval with respect to the space-time metric \cite{Mao:2006bb}. While in GR autoparallels and extremals curves do coincide and we can simply speak of geodesics,  the same is not true when torsion is present.

To infer bounds on $\alpha,\Lambda$,  we will look at the geocentric right ascension RA, declination DEC and range $\rho$, which are directly observable quantities usually measured in accurate tracking of spacecraft orbiting  some of the major bodies of the Solar System like MESSENGER (Mercury) and Cassini (Saturn). We will exploit the existing time series  $\Delta{\rm RA}_{\rm exp}(t),\Delta{\rm DEC}_{\rm exp}(t),\Delta\rho_{\rm exp}(t)$ of the post-fit residuals of such observables produced by dedicated teams of astronomers by comparing long data records with the corresponding theoretically computed observations on the basis of accurate multiparameteric models of the orbital dynamics of the probes themselves, of the planets, of the propagation of the electromagnetic waves, of the measuring devices, etc. In such data reductions, dozens of such parameters like, e.g. the planetary initial conditions, etc., are determined in a least-square sense.

Usually, only standard dynamics in terms of the currently accepted Einsteinian laws of gravitation is modeled, so that, in principle, the residuals  $\Delta{\rm RA}_{\rm exp}(t),\Delta{\rm DEC}_{\rm exp}(t),\Delta\rho_{\rm exp}(t)$, which, by construction, account for any  unmodeled and mismodeled effects, should be affected also by the modified gravity studied here. Thus, we will produce time series $\Delta{\rm RA}(t),\Delta{\rm DEC}(t),\Delta\rho(t)$ of simulated residuals due to given values of $\alpha$ and/or $\Lambda$ and compare them to the real residuals   existing in the literature to infer the limiting values of $\alpha,\Lambda$ which makes our theoretical signatures compatible with
$\Delta{\rm RA}_{\rm exp}(t),\Delta{\rm DEC}_{\rm exp}(t),\Delta\rho_{\rm exp}(t)$. Although not strictly rigorous, our method is able to yield reasonable answers about what can be inferred on the basis of the state-of-the-art of the planetary ephemerides. Indeed, if, on the one hand, the exotic dynamical models one is interested in should be explicitly modeled and dedicated solved-for parameters should be determined in ad-hoc newly generated ephemerides re-processing the same data records, on the other hand the fact that our model is relatively simple, implying just one/two extra-parameter(s), helps in assuring that the results we obtain are plausible. Moreover, it is just the case to remark that our approach, applied by a plethora of other researchers to different observation-related quantities determined for different scopes with respect to the subsequently arisen  goals of testing/constraining this or that alternative modified gravity, has always been widely followed in the literature so far in a number of scenarios, including also, e.g., the PPN parameters, the Pioneer Anomaly, etc. In some cases, like MOND \cite{Iorio_X, Cass_range} and the Pioneer Anomaly \cite{Iorio_Pio, pio1, pio2, pio3}, both approaches were applied giving substantially equivalent results.

\begin{figure*}
\centering
\centerline{
\vbox{
\begin{tabular}{cc}
\epsfysize= 5.0 cm\epsfbox{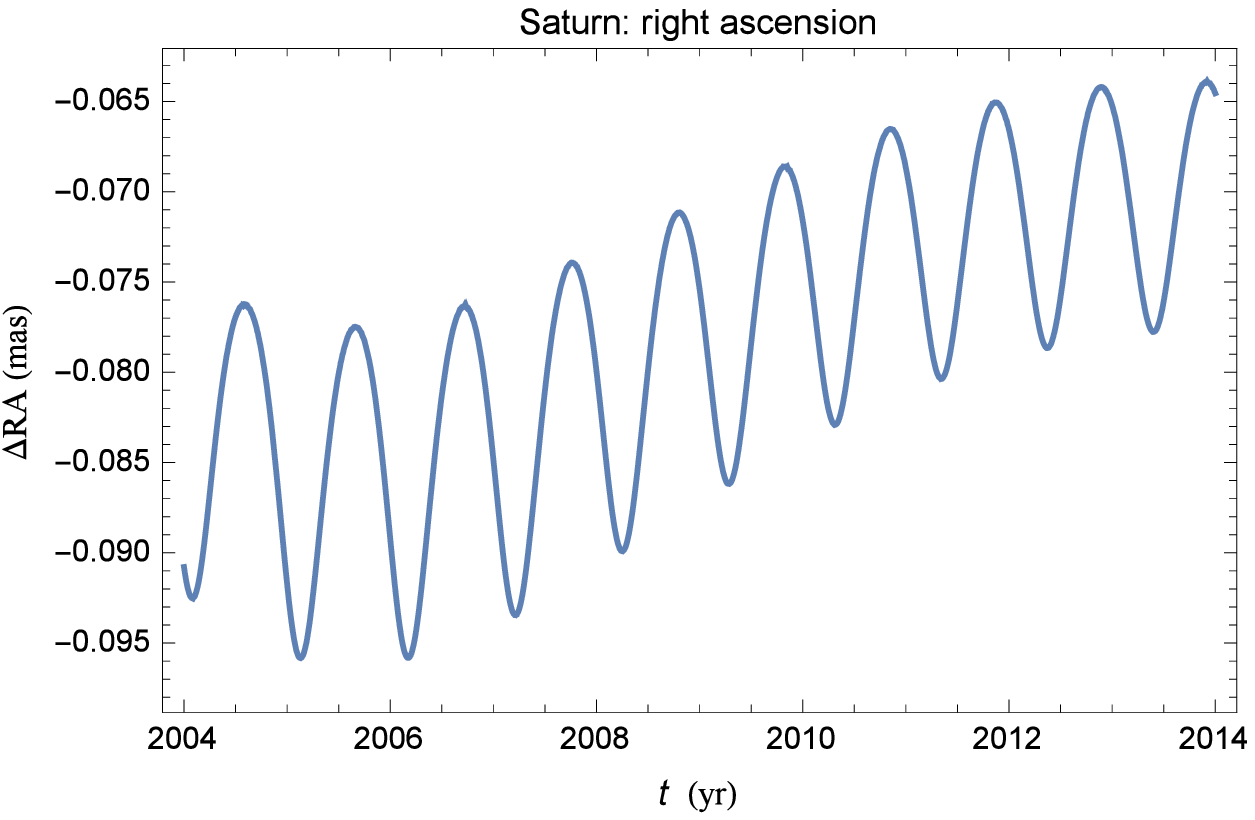} \\
\epsfysize= 5.0 cm\epsfbox{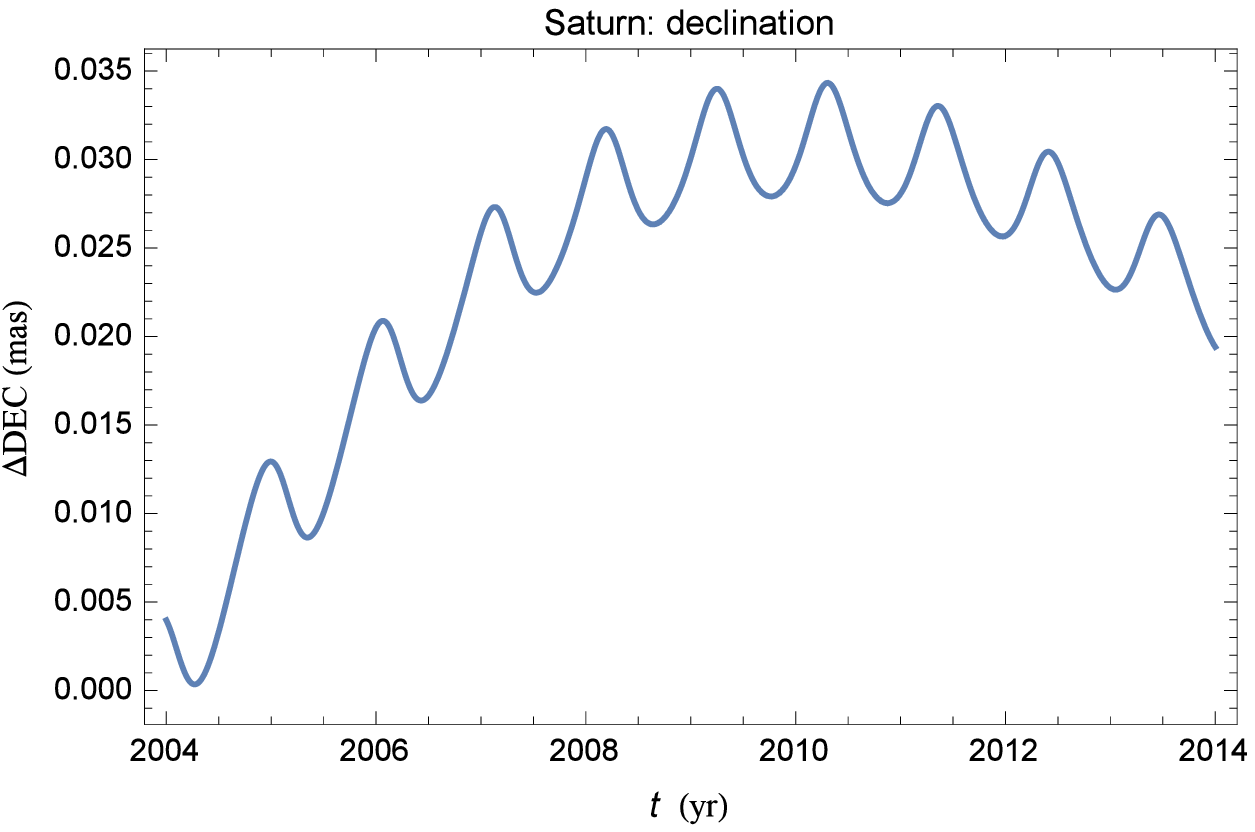} \\
\epsfysize= 5.0 cm\epsfbox{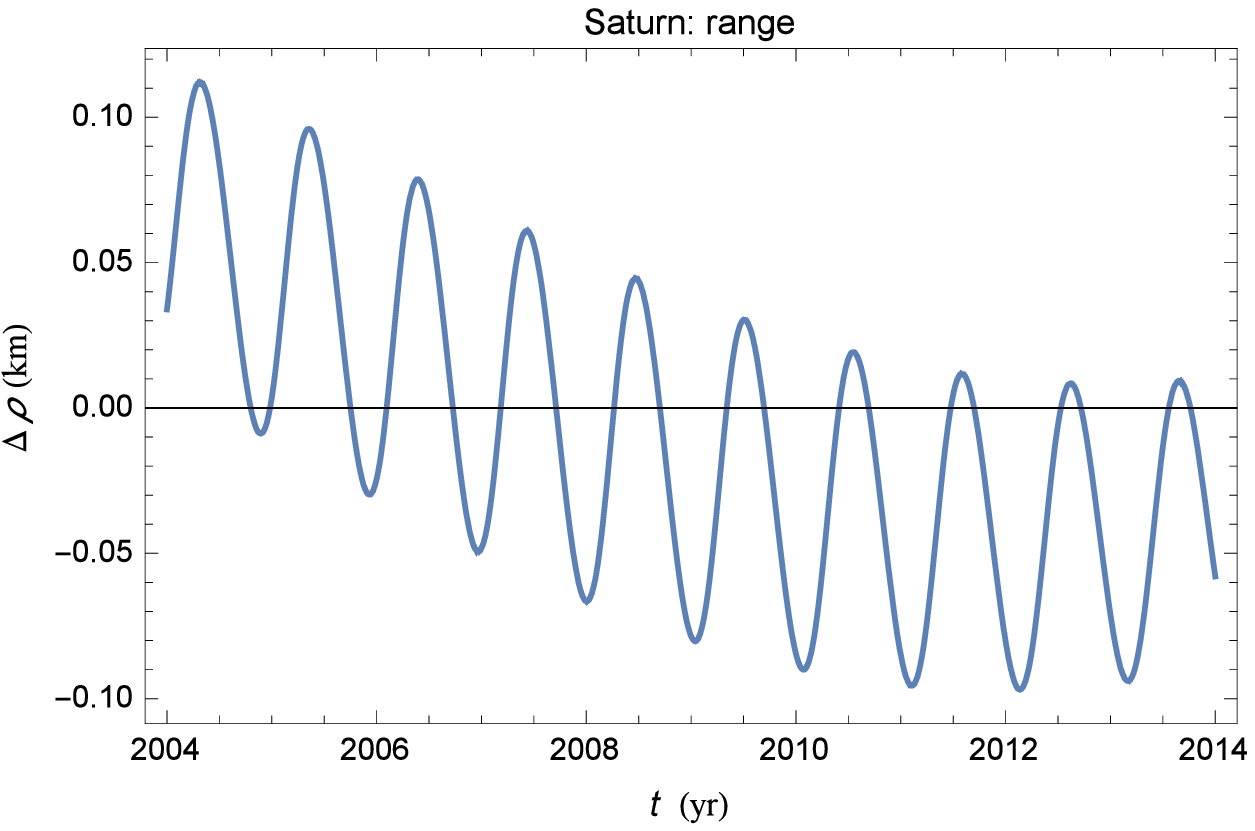} \\
\end{tabular}
}
}
\caption{Numerically produced time series of RA, DEC, $\rho$ of Saturn perturbed by $\Lambda=5 \times 10^{-44}$ m$^{-2}$. For each of the three Kronian observables considered, they were calculated as differences between two numerical integrations of the SSB barycentric equations of motion of the Sun, its eight planets and the dwarf planet Pluto from 1914 to 2014 with and without the $\Lambda$ term. Both integrations shared the same initial conditions in rectangular Cartesian coordinates, retrieved from the HORIZONS web interface, and the same standard dynamical models, apart from $\Lambda$ itself.
Thus, such curves represent the expected $\Lambda$-induced signatures  $\Delta{\rm RA},~\Delta{\rm DEC},~\Delta\rho$.
The patterns and the size of the present signals can be compared with the range residuals by \cite{Cass_range} ($\left|\Delta\rho\right|_{\rm exp}\lesssim 0.1$ km) and those for RA, DEC by \cite{Cass_VLBI} ($\left|\Delta{\rm RA}\right|_{\rm exp},\left|\Delta{\rm DEC}\right|_{\rm exp}\lesssim 4$ milliarcseconds).}\label{figura_Sat}
\end{figure*}

\begin{figure*}
\centering
\centerline{
\vbox{
\begin{tabular}{cc}
\epsfysize= 5.0 cm\epsfbox{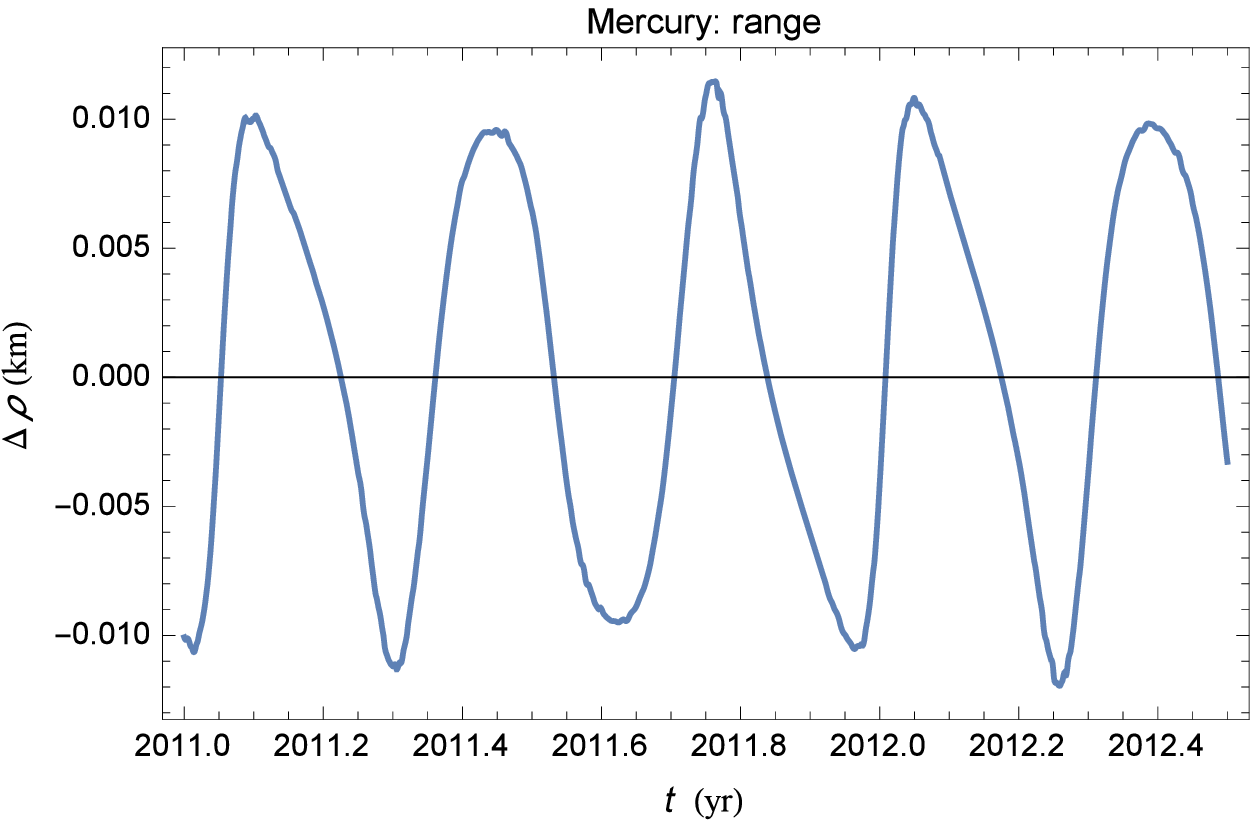} \\
\end{tabular}
}
}
\caption{Numerically produced time series of $\rho$ of Mercury perturbed by $\alpha=5 \times 10^{-1}$ m$^{2}$. It was calculated as the difference between two numerical integrations of the SSB barycentric equations of motion of the Sun, its eight planets and the dwarf planet Pluto from 1914 to 2014 with and without the $\alpha$ term. Both integrations shared the same initial conditions in rectangular Cartesian coordinates, retrieved from the HORIZONS web interface, and the same standard dynamical models, apart from $\alpha$ itself.
Thus, such curve represents the expected $\alpha$-induced signature  $\Delta\rho$.
The pattern and the size of the present signal can be compared with the range residuals by \cite{Fienga_Mess} ($\left|\Delta\rho\right|_{\rm exp}\lesssim 8.4$ m). Note that most of such a figure may be due to the unmodeled Lense-Thirring effect, which should be subtracted by leaving a smaller amplitude for the Mercury's reasiduals.}\label{figura_Mer}
\end{figure*}

In order to produce our simulated residuals, we first numerically integrate the Solar System's barycentric equations of motion of the Sun, the eight planets and the dwarf planet Pluto with and without the $\alpha/\Lambda$ additional accelerations. Both the integrations share the same initial conditions, retrieved from the HORIZONS web interface maintained by NASA JPL. Then, for a given Solar System's major body and for suitably chosen values of $\alpha$ and $\Lambda$, we take the difference between both integrations and numerically calculate the time series of its RA, DEC, range which show the characteristic pattern of the impact of the modified gravity considered on such observables: they are our theoretical templates, to be compared with the real residuals. In view of the different dependence on $r$ of the $\alpha$ and $\Lambda$ extra-accelerations, we will use  Saturn for $\Lambda$ (Figure \ref{figura_Sat}) and Mercury for $\alpha$ (Figure \ref{figura_Mer}).

The orbit of Saturn was recently constrained to a very accurate level from the analysis of long records of various types of observations from the Cassini spacecraft, engaged in the exploration of the Kronian system since 2004: its end is currently scheduled for\footnote{See http://saturn.jpl.nasa.gov/mission/saturntourdates/saturntourdates2017/ on the WEB.} late 2017, with the atmospheric entry into the planet. The geocentric range of the ringed planet is currently known to a $0.1$ km level \cite{Cass_range} from the precise radio-tracking of the spacecraft, while its RA and DEC residuals were recently reduced down to a few milliarcseconds level \cite{Cass_VLBI}  by means of the Very Long Baseline Array (VLBA) technique.
In Figure \ref{figura_Sat}, we show our simulated signatures for
\begin{equation}\Lambda=5 \times 10^{-44}\ {\rm m}^{-2}.\label{eq:boundL}\end{equation}
It turns out that the range is more effective in constraining $\Lambda$ since a larger value of it would compromise the compatibility of our $\Delta\rho(t)$ signature with the Cassini-based range residuals. The resulting bound on $\Lambda$ of (\ref{eq:boundL}) is one-two orders of magnitude tighter than those inferred, within different theoretical and phenomenological frameworks, in, e.g., \cite{IorioRug07, Iorio12, Xie:2013vua}.

As far as Mercury is concerned, its trajectory is nowadays mainly determined by range measurements to the MESSENGER spacecraft after its orbit insertion in early 2011; the mission ended this year, with a guided crash of the probe into the planet's surface.  Figure 5 of \cite{Mess_range} depicts the range residuals covering about 6 months produced at JPL, NASA with a recent version of the DE ephemerides; they appear to be far smaller than 50 m. In \cite{Fienga_Mess}, an independent team from IMCCE processed one and half years of radioscience data of MESSENGER with the INPOP13a ephemerides by constraining the geocentric range residuals of Mercury down to  $\left|\Delta\rho\right|_{\rm exp}\leq 8.4$ m. In Figure \ref{figura_Mer}, our $\alpha$-induced simulated range residuals obtained for \begin{equation}\alpha = 5\times 10^{-1}\ {\rm m}^2\label{eq:bounda}\end{equation} are shown; a larger value for $\alpha$ would make them incompatible with the existing MESSENEGR-based real residuals in \cite{Fienga_Mess}.
The bound of (\ref{eq:bounda}) is about one-three orders of magnitude better than those in \cite{Iorio12, Xie:2013vua,Ruggiero:2015oka}. As remarked, e.g., in \cite{Xie:2013vua}, actually, it could be even tighter if a smaller value for $|\Delta \rho|_{\rm exp}$ was to be used. Indeed, neither in \cite{Mess_range} nor in \cite{Fienga_Mess} the Solar gravitomagnetic field was modeled, so that it is likely that most of the Hermean residuals produced so far are due to such general relativistic dynamical effect, which should be subtracted by leaving an even smaller unmodelled/mismodeled signature. However, we conservatively assume (\ref{eq:bounda}).

\section{Summary and Conclusions}\label{sec:conc}

In this paper, we used the latest results from Solar System planetary ephemerides to study the impact of $f(T)$ gravity. In particular, we focused on spherically symmetric solutions obtained for Lagrangians in the form  $f(T)=T+\alpha T^{n}$. Since the case $f(T)=T$ corresponds to TEGR gravity, which is equivalent to GR, $\alpha$ is a  constant that parameterizes the departure of these theories from GR, hence it is expected to be small. Due to the lack of  Lorentz invariance for local transformations, we used a non diagonal tetrad to obtain the solutions for the field equations and we also considered the presence of a cosmological constant $\Lambda$.

\textcolor{black}{The presence of the cosmological constant naturally suggests the comparison of the constraints obtained in this context with those coming from cosmological solutions. Indeed, an alternative model of gravity should be phenomenologically viable and congruent in different settings and at least reproduce the amazing coherence of General Relativity.
For the $f(T)$ model a complete study of exact cosmological solutions coming from power law model via the Noether symmetry approach is presented in \cite{Basilakos:2013rua}, while the comparison with cosmological data has been reported in \cite{cardone12} and the imprints on galaxy clustering and weak gravitational lensing are derived in \cite{Camera:2013bwa}.
Nevertheless, a direct confrontation of the results is not possible, as it can be deduced by a rapid inspection of the torsion scalar, that is negative definite for the diagonal tetrads suitable for the solutions in the cosmological setting ( with no spatial curvature).
The correct way to proceed would be to derive the junction conditions for the solutions describing the different regions of spacetime one wants to match, to make sure of their accomplishment and then to infer something on the parameters. At the present stage the junction conditions for the $f(T)$ model have been derived in \cite{delaCruz-Dombriz:2014zaa}, and they turn out to be more restrictive than in GR since they depend both on the detail of the model considered and on the choice of the tetrads. Actually, it plays a crucial role due to the fact that the induced tetrads on the boundary have to be continuous; we are going to investigate this issue in future works.}

We used the aforementioned spherically symmetric solutions to describe the Sun's gravitational field and  used the Solar System data to constrain the model parameters $\alpha$ and $\Lambda$, for the case $n=2$, since for $n>2$ the impact of the non linearity of the Lagrangian is expected to be smaller. To this end, we considered the  right ascension, declination and range measured for the spacecrafts MESSENGER and Cassini, orbiting Mercury and Saturn, respectively. Our approach is based on the employment of the time series of the post-fit residuals of these observables:  in fact, since in the models of orbital dynamics only standard GR is taken into account, deviations from GR should affect the residuals. So, we calculated time series of simulated residuals due to given values of $\alpha$ and $\Lambda$, and then we compared them to the actual values of the residuals to obtain the limiting values of these parameters that are compatible with the observations.  This kind of approach, even though is not strictly rigorous, can be anyway used to infer plausible results, as we discussed in Section \ref{sec:bound}.

As a result, the bounds that we obtained  both for the cosmological constant $\Lambda$ and for the parameter $\alpha$, are tighter than those available in literature deriving by Solar System dynamics. In particular, for  the cosmological constant the bound is $\Lambda=5 \times 10^{-44}\ {\rm m}^{-2}$, while for the $f(T)$ Lagrangian parameter, the bound is $\alpha = 5\times 10^{-1}\ {\rm m}^2$.

\end{document}